\documentclass{PoS}
\usepackage{amssymb}
\usepackage{graphicx}
\usepackage{amsfonts, color, float, bbm, amsmath}

\def\amulohvp{$a_\mu^{LO,HVP}$}
\def\hatamulohvp{$\hat{a}_\mu^{LO,HVP}$}

\title{A new strategy for evaluating the LO HVP contribution to $(g-2)_\mu$ 
on the lattice}

\ShortTitle{A new strategy for evaluating the LO HVP contribution to 
$(g-2)_\mu$ on the lattice}

\author{\speaker{K. Maltman}%
\\
        York University and CSSM, University of Adelaide \\
        E-mail: \email{kmaltman@yorku.ca}}

\author{M. Golterman\\
        San Francisco State University\\
        E-mail: \email{maarten@stars.sfsu.edu}}

\author{S. Peris\\
        Universitat Aut\`onoma de Barcelona\\
        E-mail: \email{peris@ifae.es}}

\abstract{A highly physical model of the subtracted $I=1$ vector 
polarization, obtained using a dispersive representation
with precise hadronic $\tau$ decay data as input, is used to investigate 
systematic issues in the lattice evaluation of the 
leading order hadronic vacuum polarization contribution to the anomalous 
magnetic moment of the muon. The model is also employed to study 
possible resolutions of these problems. A hybrid approach to 
analyzing lattice data, involving low-order Pad\'e, low-degree 
conformal-variable polynomial, or supplemented NNLO ChPT fits 
for $Q^2$ below $\sim 0.1-0.2$ GeV$^2$ 
and direct numerical integration of lattice data
above this point, is shown to bring the systematic issues identified
under control at the sub-$1\%$ level.}

\FullConference{The 32nd International Symposium on Lattice Field Theory\\
		 23-28 June, 2014\\
		 Columbia University New York, NY}

\begin{document}

\section{\label{intro}Introduction}
The muon anomalous magnetic moment, $a_\mu\equiv (g-2)_\mu /2$, is 
currently measured to $0.5$ ppm~\cite{bnlgminus2}, with plans for a 
further factor of $4$ improvement in the upcoming Fermilab experiment. 
This improvement is of particular interest given the $3.3-3.6\, \sigma$ 
discrepancy between the current determination and the Standard Model 
(SM) prediction~\cite{SMgminus2ref}. After the purely QED 
contribution~\cite{kinoshita5loopqed}, the next largest SM contribution 
is that from the leading order (LO) hadronic vacuum 
polarization (HVP), \amulohvp. The SM version of \amulohvp  is
given by the weighted dispersive integral 
\begin{equation}
\left({\frac{m_\mu^2}{12\pi^3}}\right)\, \int_{m_\pi^2}^\infty
ds\, {\frac{K(s)}{s}}\, \sigma^0_{had}(s)\, ,
\label{amulohvpdisp}\end{equation}
with $\sigma^0_{had}(s)$ the bare $e^+e^-\rightarrow\, hadrons$ 
cross-section and $K(s)$ a known kernel varying monotonically from
$0.4$ to $1$ as $s$ varies from $m_\pi^2$ to $\infty$.
The error on \amulohvp  is the largest of the errors 
entering the SM prediction~\cite{SMgminus2ref} and hence a key 
target for near-term improvement. Discrepancies between the results 
of different experiments for the key $e^+ e^-\rightarrow \pi^+ \pi^-$ 
cross-sections~\cite{sigpipi} also represent an important 
complication for the dispersive evaluation.

The deviation of the experimental result from the SM 
prediction, and the role played by the error on 
\amulohvp, have led to interest in an independent 
determination of \amulohvp  from the 
lattice~\cite{blum03,latticeamulohvp,abgp12,gmpamu1,hpqcd14,gmpamu2}.
Such a determination is made possible by the alternate 
representation~\cite{blum03,lpr72}
\begin{equation}
a_\mu^{LO,HVP}=\, -4\alpha_{EM}^2\, \int_0^\infty dQ^2\, f\left( Q^2\right)\,
\hat{\Pi}\left( Q^2\right)\, ,
\label{amulohvplatt}\end{equation}
with the integral running over Euclidean $Q^2$,
$\hat{\Pi}(Q^2)=\Pi (Q^2)-\Pi (0)$ the subtracted
electromagnetic (EM) current polarization and $f\left( Q^2\right)$ a 
known kernel which makes the integrand very strongly peaked at low $Q^2$
($Q^2\simeq m_\mu^2/4$). 

The lattice evaluation is complicated by the fact that, for current 
simulations, the discrete $Q^2$ available on the lattice provide rather 
coarse coverage of the critical low-$Q^2$ region, with the lowest 
accessible $Q^2$ lying above the peak of the integrand in 
Eq.~(\ref{amulohvplatt}) and the errors at the lowest few $Q^2$ points 
typically large (see, {\it e.g.}, Fig.~\ref{fig1b}). These limitations 
make an accurate determination of \amulohvp  by direct numerical 
integration of lattice data impossible at present. The problem is dealt 
with by fitting a continuous form to $\hat{\Pi}(Q^2)$ and using the 
resulting fitted version in place of $\hat{\Pi}(Q^2)$ in 
Eq.~(\ref{amulohvplatt}). Existing lattice analyses 
have typically employed fit ranges extending up to $Q^2\sim 1$ 
or $2$ GeV$^2$, the much smaller errors at larger $Q^2$
serving to sharpen the determinations of the parameters employed in 
the fits. The very good $\chi^2/dof$'s typically obtained, however,
provide no information about the reliability of the extrapolation 
of the fits to the very low-$Q^2$ region most relevant 
to $a_\mu^{LO,HVP}$. The reason is as follows.
With $\rho (s)\ge 0$ the EM current spectral function, $\hat{\Pi}(Q^2)$ 
satisfies the dispersion relation
\begin{equation}
\hat{\Pi}(Q^2)=\,-Q^2\, \int_{th}^\infty ds\;
\frac{\rho (s)}{s(s+Q^2)}\, ,
\label{dispreln}\end{equation}
with $th$ the threshold. The magnitudes of 
derivatives of all orders of $\hat{\Pi}(Q^2)$ with respect to $Q^2$
\begin{equation}
\vert d^n\, \hat{\Pi}(Q^2)/(dQ^2)^n \vert
=\, n! \, \int_{th}^\infty ds\ \rho (s)/(s+Q^2)^{n+1}\, ,
\label{derivsdispreln}\end{equation}
are thus uniformly smaller at high $Q^2$ than low $Q^2$. Fits to 
$\hat{\Pi}(Q^2)$ over a wide range of $Q^2$ therefore suffer from a
potential systematic bias in which the much more numerous low-error, 
high-$Q^2$ points, which typically dominate the fit, produce
an underestimate of the curvature of $\hat{\Pi}(Q^2)$ at low $Q^2$
and hence an unreliable extrapolation into the very low-$Q^2$
region. This issue has been investigated in 
Refs.~\cite{gmpamu1,gmpamu2}, using a highly physical model of the 
flavor $ud$ $I=1$ vector (V) current analogue, $\hat{\Pi}^{I=1}(Q^2)$, 
of $\hat{\Pi}(Q^2)$. The results of this investigation are outlined 
in Sec.~\ref{sec2}.

The spectral function, $\rho^{I=1}(s)$, of $\hat{\Pi}^{I=1}(Q^2)$, 
appearing in the $I=1$ analogue of Eq.~(\ref{dispreln}), is experimentally 
determinable for $s< m_\tau^2$ from the inclusive $ud$ V hadronic $\tau$ 
decay distribution~\cite{tsai}, and modelled for $s>m_\tau^2$ as a sum 
of 5-loop $D=0$ OPE~\cite{5loopPT} and residual duality-violating (DV) 
contributions, the latter treated using a large-$N_c$ and Regge motivated 
model with parameters obtained in the finite energy sum rule analyses of 
Ref.~\cite{boitoetalalphasdv}. The region $s>m_\tau^2$ plays a very 
small role at the low $Q^2$ relevant to $a_\mu^{LO,HVP}$, making the 
resulting $\hat{\Pi}^{I=1}(Q^2)$ model a highly physical one for use in 
quantitative studies of the systematics of fit approaches employed in 
earlier lattice analyses.

In what follows, \hatamulohvp  denotes the $ud$ V analogue 
of \amulohvp, $\hat{a}_\mu^{LO,HVP}[Q^2_{min},Q^2_{max}]$ the 
contribution from the interval $Q^2_{min}\le Q^2\le Q^2_{max}$ 
in the analogue of Eq.~(\ref{amulohvplatt}), and 
$\hat{a}_\mu^{LO,HVP}[Q^2_{max}]\equiv\hat{a}_\mu^{LO,HVP}[0,Q^2_{max}]$.
We also employ a set of fake $ud$ V ``lattice data'' obtained by drawing 
a random sample from a multivariate Gaussian distribution generated using 
central values from the dispersive model for $\hat{\Pi}^{I=1}(Q^2)$ and 
the $Q^2$ set and covariance matrix of a $64^3\times 144$ MILC ensemble 
with periodic boundary conditions, lattice spacing $a\approx 0.06$~fm 
and $m_\pi\approx 220$~MeV~\cite{MILC}.

\section{\label{sec2}The systematic problem and a hybrid strategy
for its solution}
In what follows, we investigate the systematic issues raised above
using the dispersive $I=1$ model and the fake lattice data generated
from it. To make the impact on the accuracy with which \hatamulohvp  
can be evaluated as transparent as possible, we will quote all errors 
as fractions of the full LO HVP contribution, \hatamulohvp. We assume 
that $\Pi (0)$, needed to construct $\hat{\Pi}(Q^2)$ from the computed 
lattice $\Pi (Q^2)$, can be determined with sufficient accuracy, 
referring the reader to Ref.~\cite{gmpamu2} for further discussion.
\vskip 0.12in
\begin{figure}[htb]
{\begin{minipage}[htb]{0.42\linewidth}
{\rotatebox{270}{\mbox{
\includegraphics[width=1.8in]{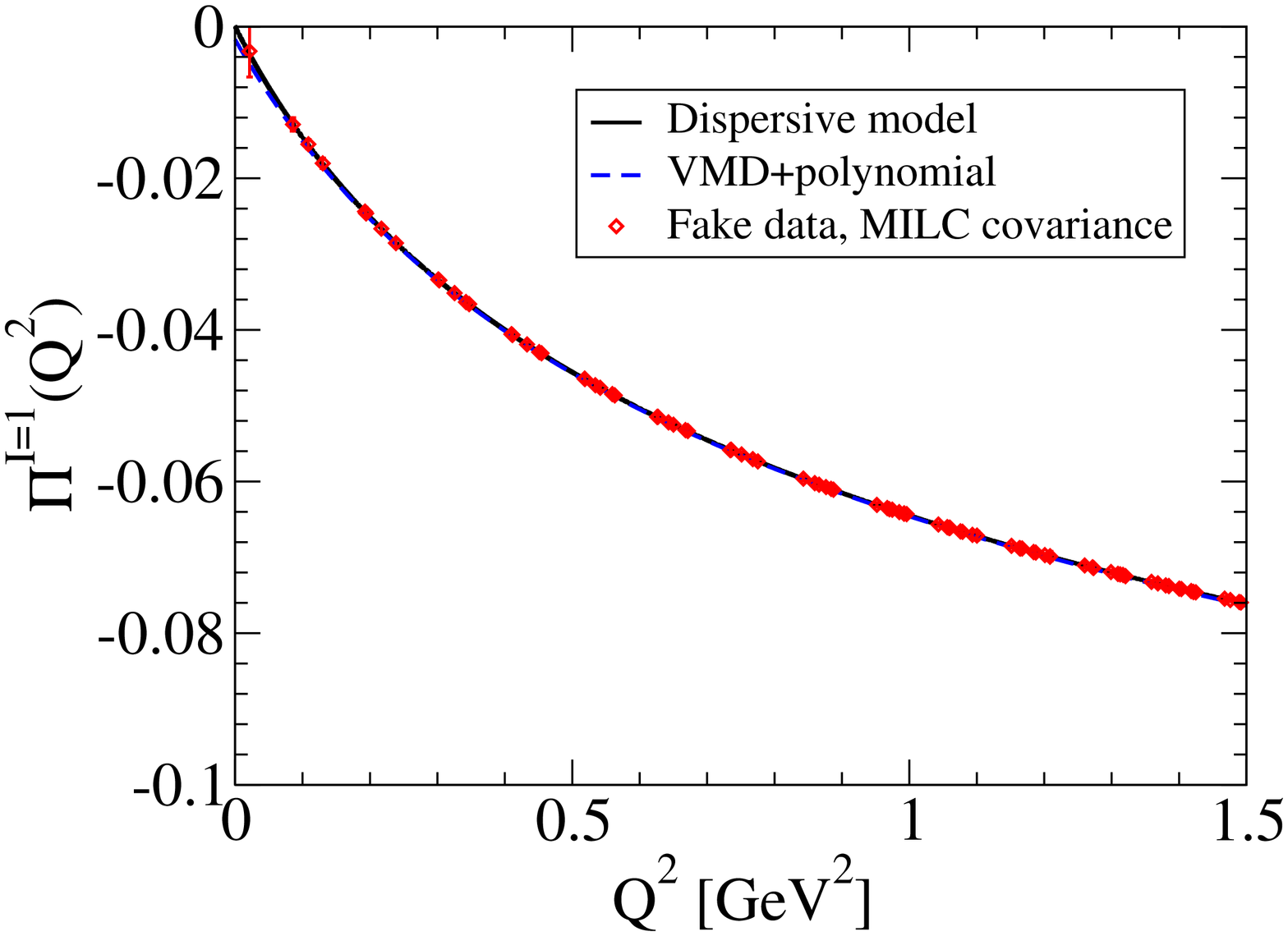}
}}}
\caption{\label{fig1a} VMD+ fit (blue dashed curve),
fake lattice data (red points) and underlying dispersive
model (black solid curve) for $\hat{\Pi}^{I=1}(Q^2)$,
$0<Q^2\le 1$~GeV$^2$}
\end{minipage}
\qquad 
\begin{minipage}[htb]{0.42\linewidth}
{\rotatebox{270}{\mbox{
\includegraphics[width=1.8in]{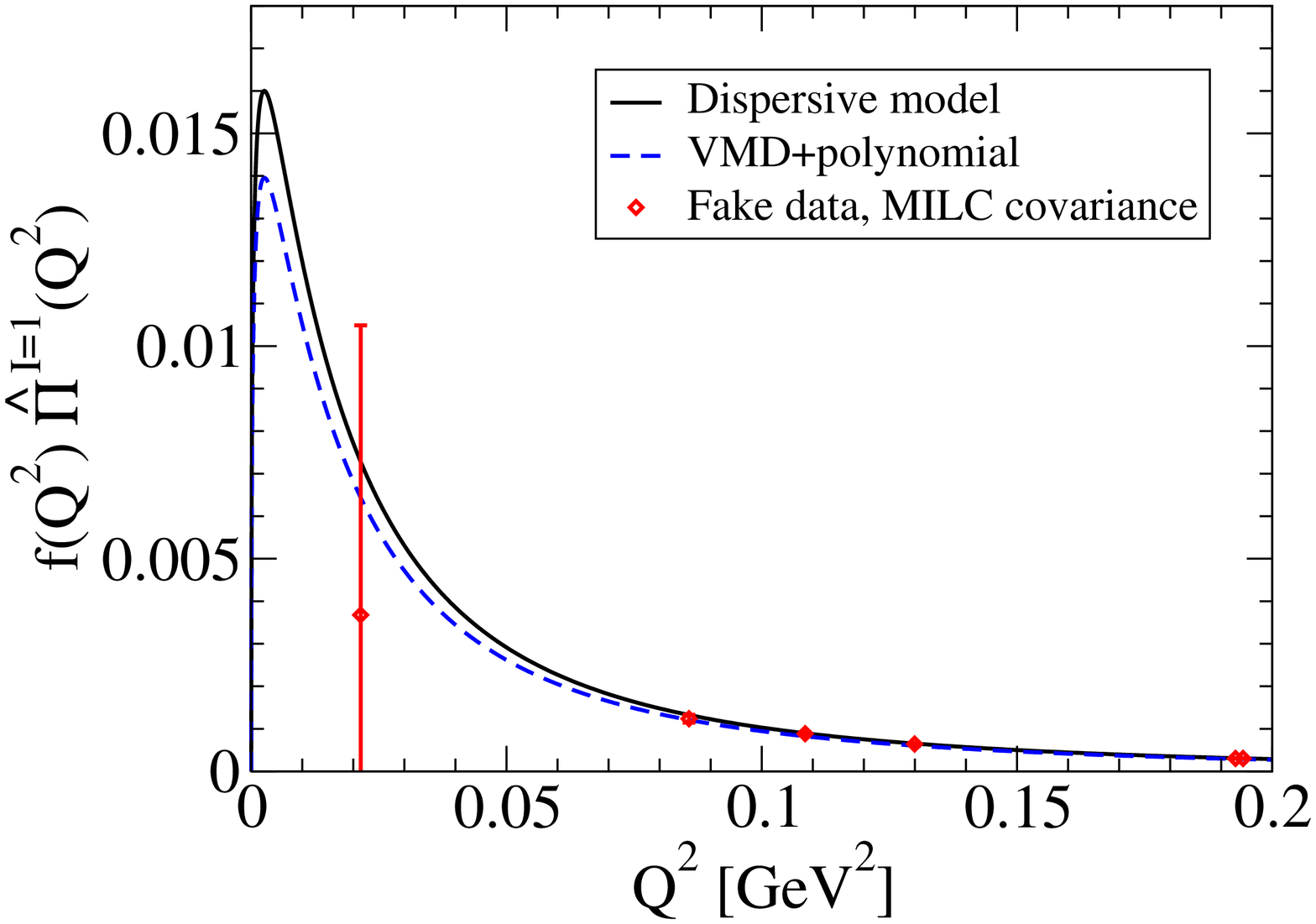}
}}}
\caption{\label{fig1b} VMD+ fit (blue dashed curve),
fake lattice data (red points) and underlying dispersive
model (black solid curve) for $f(Q^2)\, \hat{\Pi}^{I=1}(Q^2)$,
$0<Q^2\le 0.2$~GeV$^2$}
\end{minipage}}
\end{figure}

Fig.~\ref{fig1a} shows the result of a fit to the fake lattice data 
on the interval $0<Q^2<1$ GeV$^2$ using a form, ``VMD+'' (VMD 
plus a linear polynomial in $Q^2$), used previously in the 
literature. The quality of the fit is evidently very good. However, if 
we investigate how safe it is to extrapolate the fit into the
region of the peak of the integrand in the $I=1$ analogue of
Eq.~(\ref{amulohvplatt}), we find the result shown in 
Fig.~\ref{fig1b}~\cite{gmpamu1}. The fit significantly underestimates 
the integrand in the region of the peak and produces an estimate for 
\hatamulohvp  $\sim 11\%$ lower than the exact underlying model 
value~\cite{gmpamu1}. Even worse, the ``pull'' for this fit (the 
deviation of the fit estimate from the underlying model value, in 
units of the propagated fit error) is $18$, making it clear that 
the in-principle systematic problem identified in Sec.~\ref{intro} 
is, in general, numerically very significant. The level of the 
problem depends on the fit form employed. For example, employing
instead the $[1,2]$ version of the Pad\'es advocated in Ref.~\cite{abgp12} 
on the same interval, one obtains a pull of $0.5$~\cite{gmpamu1}. The 
potential systematic bias is, however, still present: if one fits the same 
$[1,2]$ Pad\'e on the slightly enlarged interval $0<Q^2<1.5$ GeV$^2$, 
and hence includes additional low-error, high-$Q^2$ points, the pull 
increases to $4$~\cite{gmpamu1}. While the dispersive model
can, of course, be used to quantify the systematic error associated
with the use of any fit form choice, developing a strategy that 
focusses the use of fitting as much as possible on the low-$Q^2$ region 
which dominates \amulohvp  obviously represents a more attractive option.
\vskip .12in
\begin{figure}[htb]
\begin{minipage}[htb]{0.44\linewidth}
{\rotatebox{270}{\mbox{
\includegraphics[width=1.8in]
{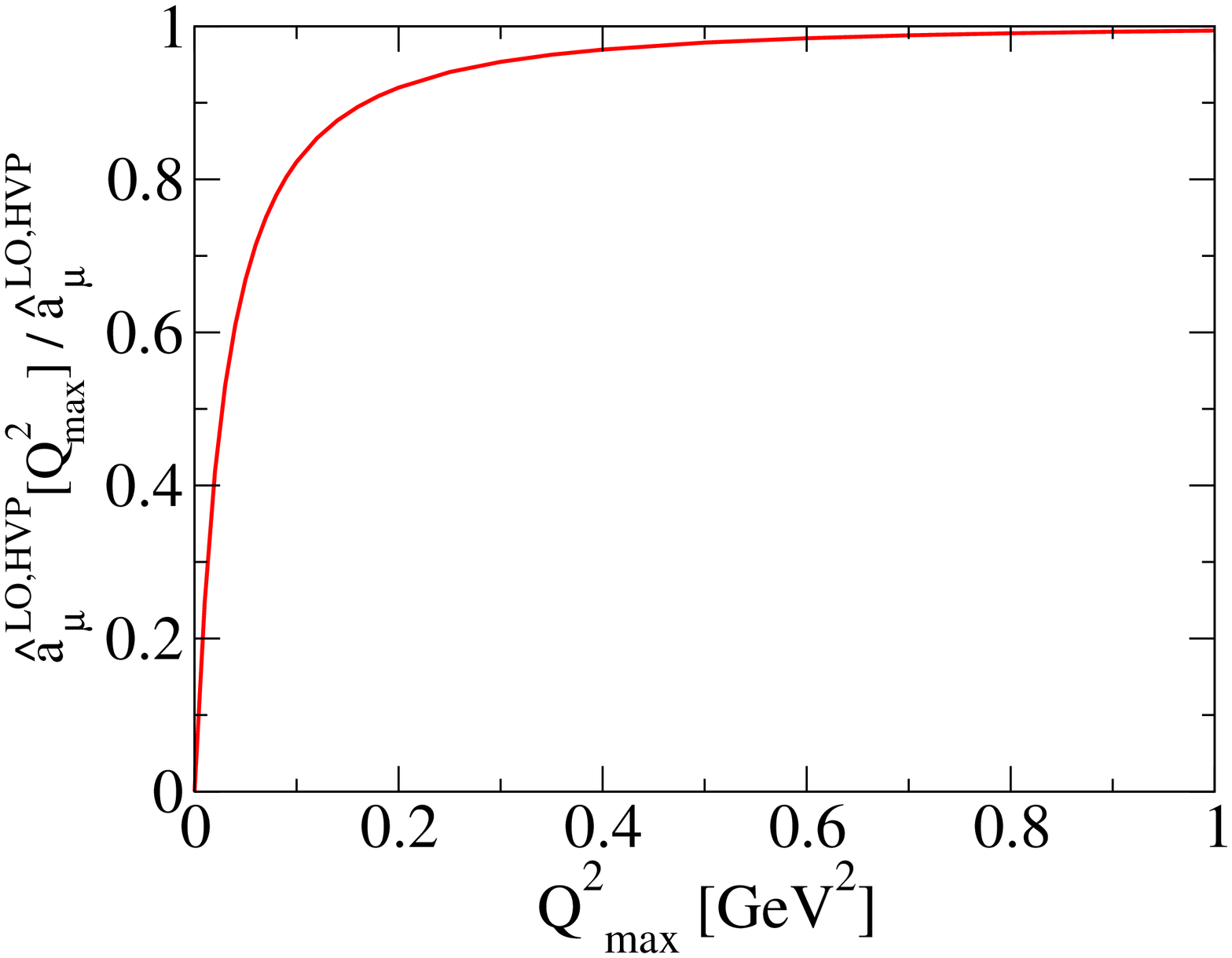}
}}}
\caption{\label{fig2}The accumulation of $\hat{a}_\mu^{\rm LO,HVP}$
as a function of $Q^2_{max}$ }
\end{minipage}
\qquad 
\begin{minipage}[htb]{0.44\linewidth}
{\rotatebox{270}{\mbox{
\includegraphics[width=1.8in]
{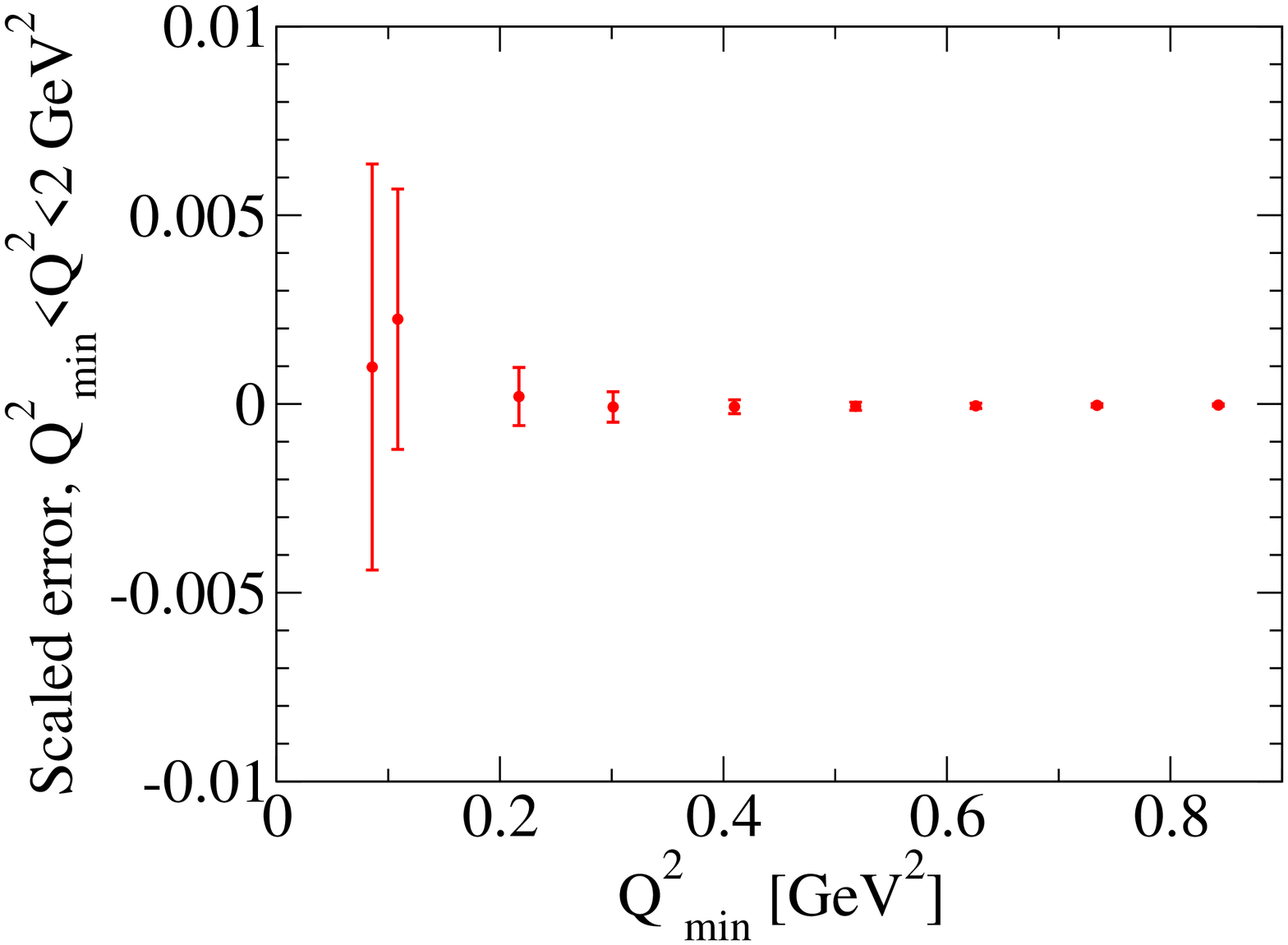}
}}}
\caption{\label{fig3}Systematic and statistical errors
on the trapezoid-rule evaluation of
$\hat{a}_\mu^{\rm LO,HVP}[Q^2_{min},2\ {\rm GeV}^2]$}
\end{minipage}
\end{figure}

Fig.~\ref{fig2} shows the accumulation of the fractional partial 
contribution, $\hat{a}_\mu^{LO,HVP}[Q^2_{max}] /\hat{a}_\mu^{LO,HVP}$,
as a function of $Q^2_{max}$. We see that over $80\%$ ($90\%$) of the 
total is accumulated below $Q^2_{max}=0.1$ ($0.2$) GeV$^2$. The 
low-$Q^2$ domination of the Euclidean integral is thus much stronger
than the low-$s$ domination of the dispersive integral, 
Eq.~(\ref{amulohvpdisp}), where one would have to integrate to
$\sim 1$ ($\sim 1.5$) GeV$^2$ to reach $80\%$ ($90\%$) of the total. 
This is a useful feature of the lattice formulation since one expects 
$\hat{\Pi}(Q^2)$ to be very accurately representable with few-parameter 
forms in such a small $Q^2$ interval. It also shows that considerably 
lower accuracy can be tolerated for the small contributions from $Q^2$ 
above $0.1-0.2$ GeV$^2$ than for the much larger low-$Q^2$ ones. It 
thus becomes relevant to investigate the accuracy with which 
higher-$Q^2$ contributions can be evaluated via direct numerical 
integration of existing lattice data. The results of this investigation, 
performed using the fake lattice data, are presented in Fig.~\ref{fig3}, 
the central values showing the systematic error, resulting from the use 
of the trapezoid-rule approximation for
$\hat{a}_\mu^{LO,HVP}[Q^2_{min},2\ {\rm GeV}^2]$, and the error 
bars the corresponding propagated statistical errors (the covariance 
matrix of the fake data is, by construction, the MILC covariance 
matrix used to generate the data). Direct numerical integration is 
found to provide a determination of the $Q^2>Q^2_{min}$ contribution 
accurate to well below $1\%$ of \hatamulohvp  for $Q^2_{min}$ 
down to $\sim 0.1$ GeV$^2$~\cite{gmpamu2}.

In light of this observation, we focus on strategies for reliably
representing the subtracted polarization in the region below
$Q^2\sim 0.1-0.2$ GeV$^2$. We have identified three approaches
capable of producing determinations of $\hat{a}_\mu^{LO,HVP}[Q^2_{max}]$
with an accuracy below $1\%$ of \hatamulohvp~\cite{gmpamu2}. The first
uses low-order Pad\'es, the second low-degree polynomials in the conformal
variable 
\begin{equation}
w(Q^2)=\left[ 1-\sqrt{1+z(Q^2)}\right] /\left[ 1+\sqrt{1+z(Q^2)}\right],
\quad z(Q^2)={\frac{Q^2}{4m_\pi^2}}\, ,
\label{confdefn}\end{equation}
and the third a supplemented form of NNLO ChPT. We discuss each of these
briefly in turn. 

\begin{figure}[htb]
\begin{minipage}[t]{0.42\linewidth}
{\rotatebox{270}{\mbox{
\includegraphics[width=1.8in]{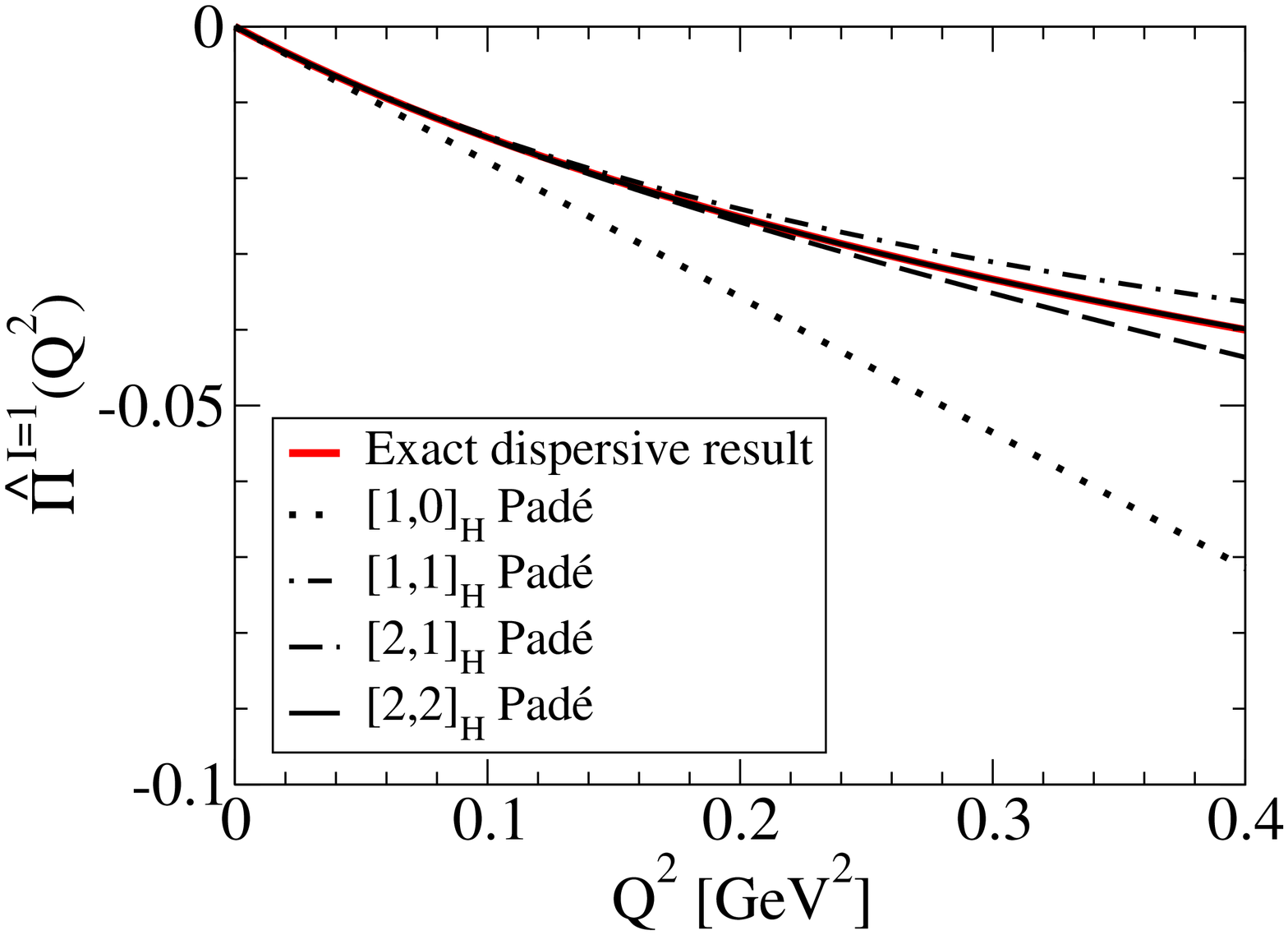}
}}}\
\caption{\label{fig4}Low-$Q^2$ comparison of one-point Pad\'e representations 
and the dispersive model for $\hat{\Pi}^{I=1}(Q^2)$}
\end{minipage}
\qquad
\begin{minipage}[t]{0.42\linewidth}
%
{\rotatebox{270}{\mbox{
\includegraphics[width=1.8in]
{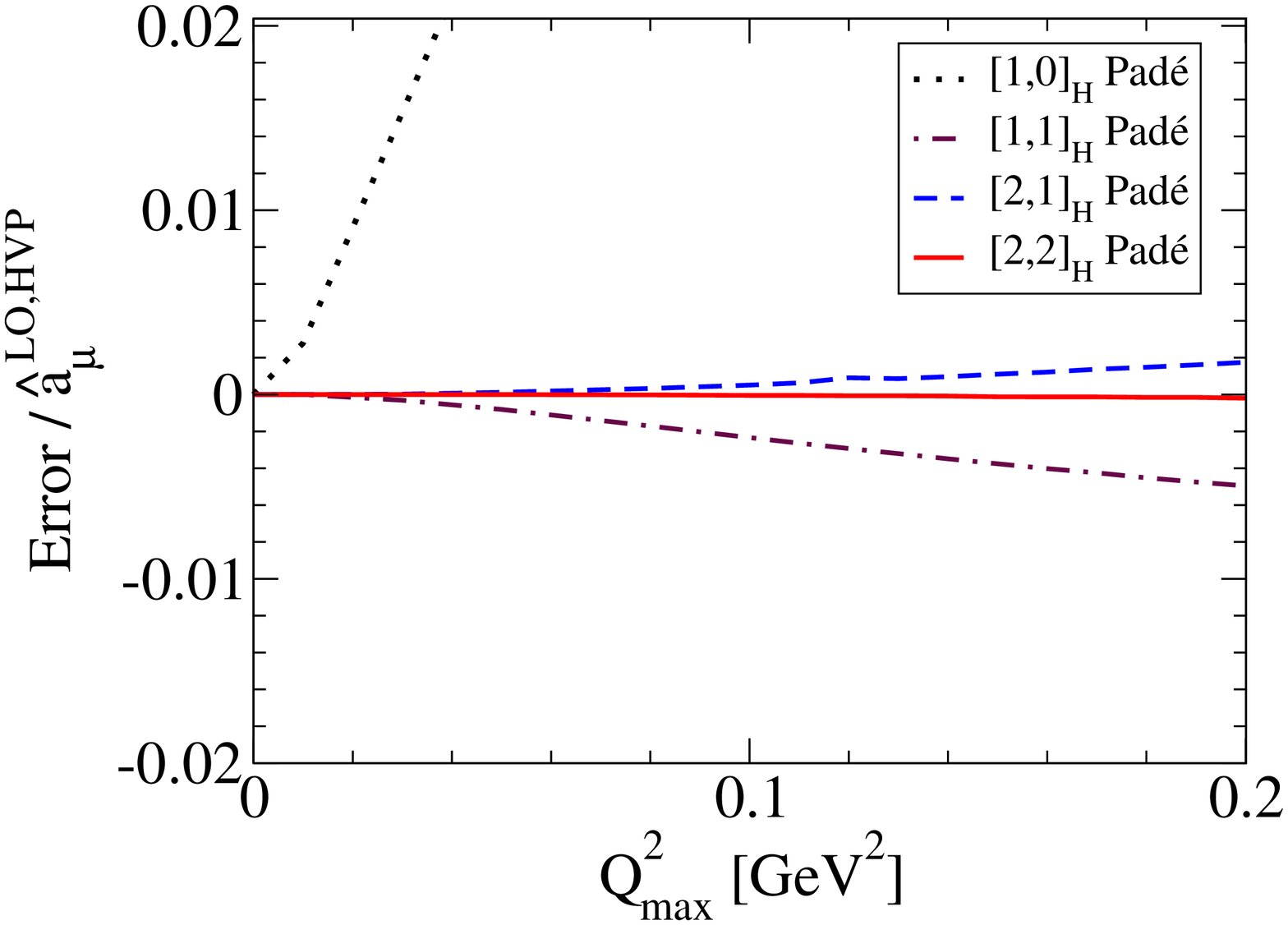}
}}}
\caption{\label{fig5}Deviations of one-point Pad\'e estimates for
$\hat{a}_\mu^{\rm LO,HVP}[Q^2_{max}] /\hat{a}_\mu^{\rm LO,HVP}$ on
the interval $0\le Q^2_{max}\le 0.2$~GeV$^2$}
\end{minipage}
\end{figure}

In the case of the Pad\'es, we first consider the ``one-point Pad\'e''
representations, constructed from the derivatives of 
$\hat{\Pi}^{I=1}(Q^2)$ with respect to $Q^2$ at $Q^2=0$. The required 
derivatives are easily determined from the dispersive representation
in the model case, and have been argued to be determinable on the lattice 
from even-order Euclidean time moments of the zero-spatial-momentum 
current-current two-point functions~\cite{hpqcd14}. In what follows, 
$[M,N]_H$ denotes the representation of $\hat{\Pi}^{I=1}(Q^2)$ as a 
quotient of polynomials of degree $M$ and $N$. Fig.~\ref{fig4} shows 
the comparison of the $[1,0]_H$, $[1,1]_H$, $[2,1]_H$ and $[2,2]_H$ 
one-point Pad\'es to the underlying dispersive model for $0<Q^2<0.4$ GeV$^2$. 
Evidently even the $[1,1]_H$ form provides an excellent representation 
below $Q^2\sim 0.2$ GeV$^2$. The corresponding errors on 
$\hat{a}_\mu^{LO,HVP}\left[ Q^2_{max}\right] /\hat{a}_\mu^{LO,HVP}$
are shown in Fig.~\ref{fig5}. The $[1,1]_H$ form yields
a result accurate to $\sim 0.3\%$ ($\sim 0.5\%$) for $Q^2_{max}=0.1$ 
($0.2$) GeV$^2$. These numbers are reduced to $\sim 0.06\%$ and 
$\sim 0.2\%$ for the $[2,1]_H$ Pad\'e. If one wished, as in 
Ref.~\cite{hpqcd14}, to use a one-point Pad\'e to evaluate 
$\hat{a}_\mu^{LO,HVP}\left[ Q^2_{max}\right]$ out to much larger 
$Q^2_{max}$, e.g., $\sim 2$ GeV$^2$, we find the $[2,2]_H$ form would 
be required to bring the systematic error down 
to the sub-percent level~\cite{gmpamu2}.
\vskip .13in

It is also possible to consider ``multi-point Pad\'es'', Pad\'es with 
coefficients obtained by fitting $\hat{\Pi}^{I=1}(Q^2)$ in an interval 
of $Q^2$. Since the low-$Q^2$ coverage is too sparse and the lowest-$Q^2$ 
errors too large to produce stable low-$Q^2$ fits of this type for the 
fake lattice data, we have investigated this possiblility using the 
dispersive $ud$ V model itself, together with the covariances generated 
by the underlying dispersive representation. The results of fits to the model 
data at $Q^2=0.10,\, 0.11,\, \cdots ,\, 0.20$ GeV$^2$ show that, to achieve 
the same accuracy as achieved with a given one-point-Pad\'e fit, it is 
necessary to go to a multi-point Pad\'e one order higher. A multi-point
$[2,1]_H$ Pad\'e, which yields an accuracy of better than $0.45\%$ on
$\hat{a}_\mu^{LO,HVP}\left[ Q^2_{max}\right] /\hat{a}_\mu^{LO,HVP}$
for $Q^2_{max}\le 0.2$ GeV$^2$, is required to reach sub-percent accuracy 
in this region for a fit of this type.
\vskip .12in
\begin{figure}[htb]
\begin{minipage}[t]{0.42\linewidth}
{\rotatebox{270}{\mbox{
\includegraphics[width=1.8in]
{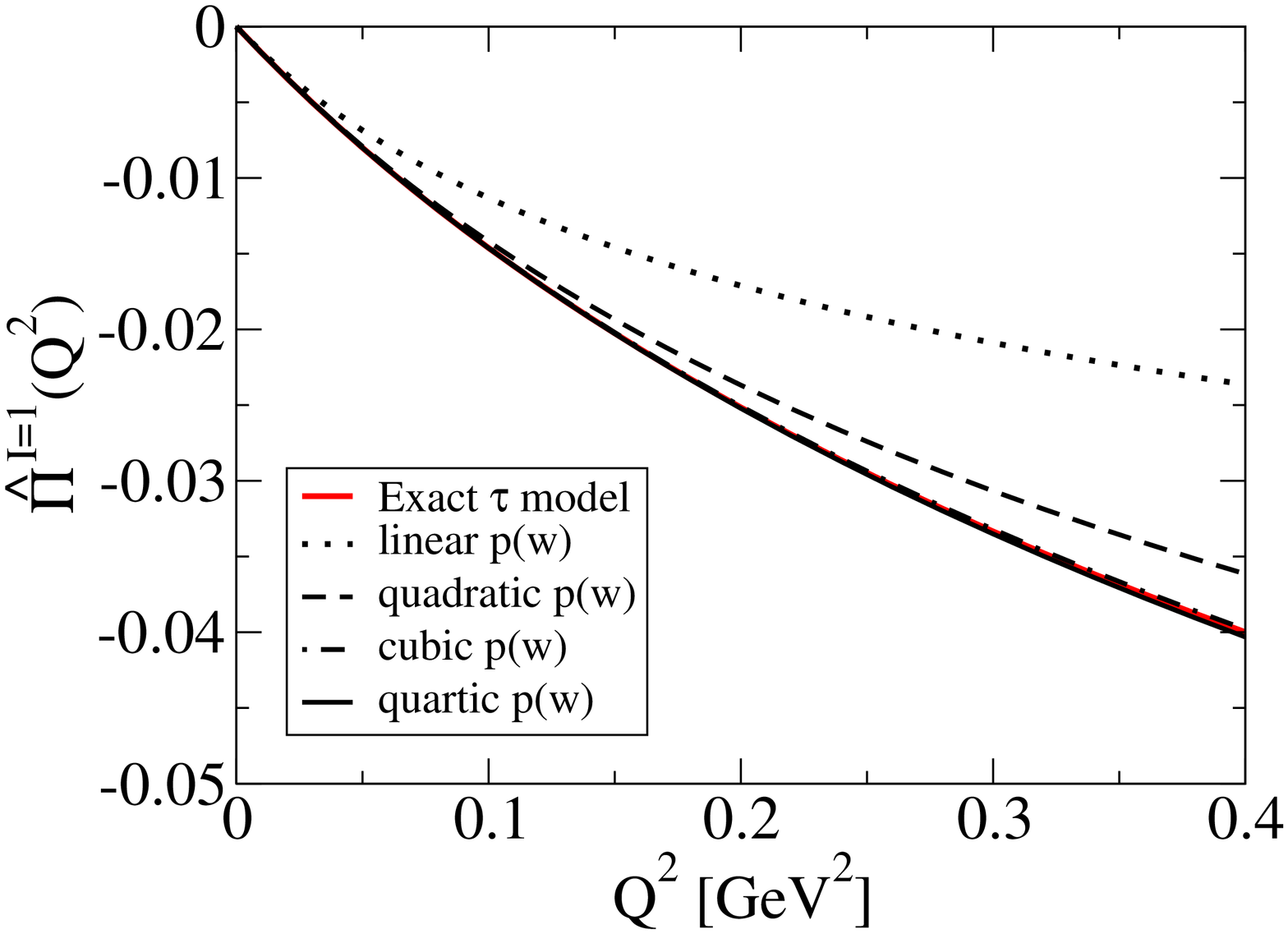}
}}}
\caption{\label{fig6}Comparison of conformal polynomial representations
to the underlying dispersive model for $\hat{\Pi}^{I=1}(Q^2)$}
\end{minipage}
\qquad
\begin{minipage}[t]{0.42\linewidth}
{\rotatebox{270}{\mbox{
\includegraphics[width=1.8in]
{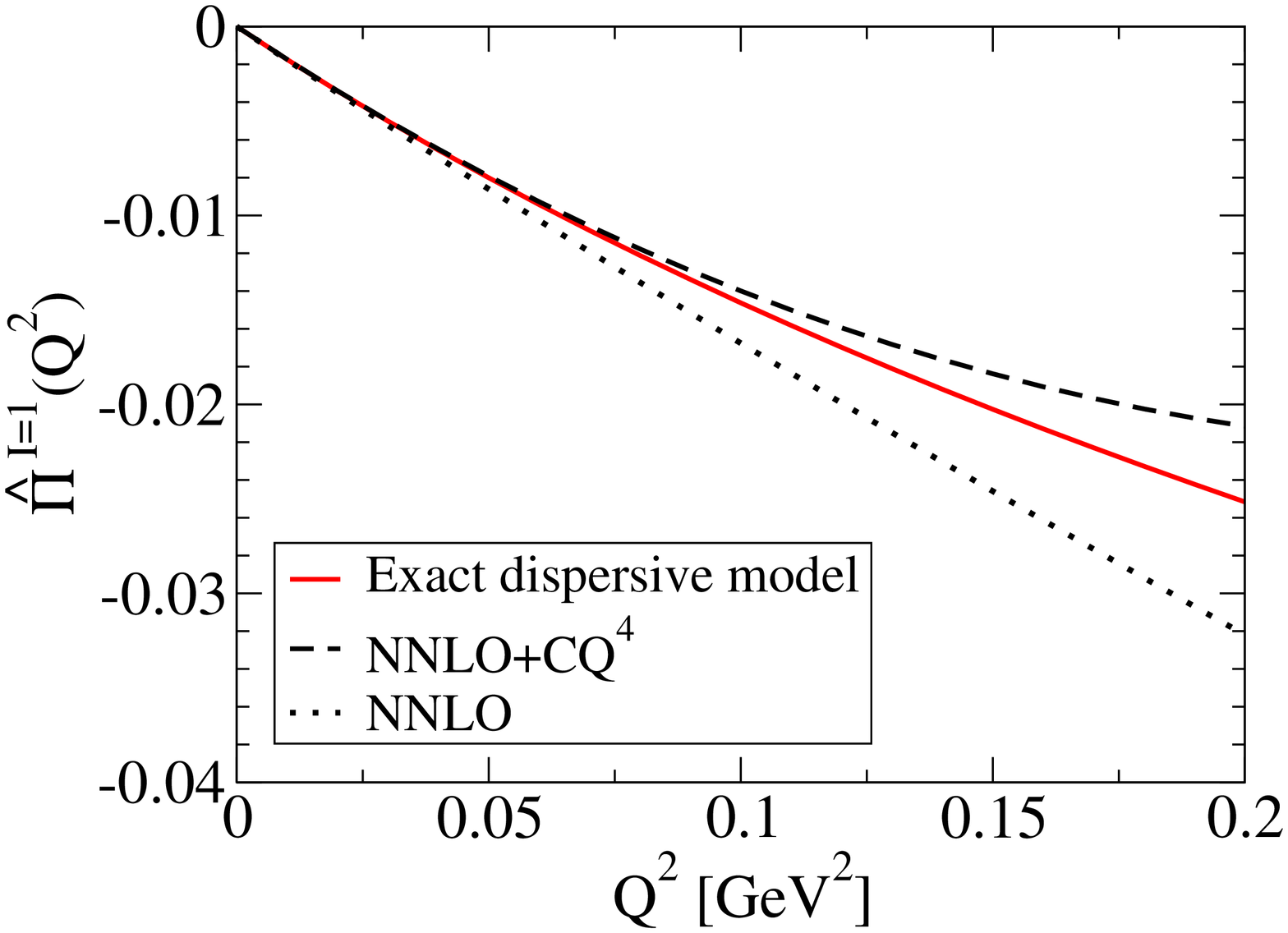}
}}}
\caption{\label{fig7}Comparison of the NN$^\prime$LO and NNLO ChPT 
representations to the underlying dispersive model for $\hat{\Pi}^{I=1}(Q^2)$}
\end{minipage}
\end{figure}

The use of polynomials in the conformal variable $w$ is motivated
by the fact that the transformation $Q^2\rightarrow w$ maps the
entire cut $Q^2$-plane into the interior of the unit circle in
the $w$-plane, with the cut mapped to the boundary. The 
Euclidean $Q^2$ entering the integral in Eq.~(\ref{amulohvplatt})
thus lie in the region of convergence of the Taylor series
in $w$. The coefficients of polynomial-in-$w$ representations of 
$\hat{\Pi}^{I=1}(Q^2)$ can, as in the Pad\'e case, be 
obtained either in ``one-point'' form from the derivatives of 
$\hat{\Pi}^{I=1}(Q^2)$ with respect to $Q^2$ at $Q^2=0$, or in 
``multi-point'' form from fits to $\hat{\Pi}^{I=1}(Q^2)$ 
on an interval of (low) $Q^2$. One-point-form results for polynomials 
linear, quadratic, cubic and quartic in $w$ are compared to the 
underlying dispersive model in Fig.~\ref{fig6}. Very accurate
representations are seen to be achievable for $Q^2$ well beyond 
$0.2$ GeV$^2$. For the quadratic case, the resulting accuracies on 
$\hat{a}_\mu^{LO,HVP}\left[ Q^2_{max}\right] /\hat{a}_\mu^{LO,HVP}$
are $0.6\%$ and $1\%$ for $Q^2_{max}=0.1$ and $0.2$ GeV$^2$, 
respectively. These numbers improve to $0.02\%$ and $0.04\%$
if the cubic representation (having the same number of parameters
as a $[2,1]_H$ Pade) is used. As in the Pad\'e
case, we find that multi-point forms one order higher must be
used to reach the same accuracy as achieved using a given one-point form
for $Q^2_{max}\le 0.2$ GeV$^2$~\cite{gmpamu2}.

The last of the low-$Q^2$ options we have considered is that provided 
by ChPT. The low-energy representation of $\hat{\Pi}^{I=1}(Q^2)$ is
known to NNLO~\cite{abt00}, and the model version of 
${\frac{d\hat{\Pi}^{I=1}}{dQ^2}}(0)$ corresponds to a value of the 
relevant NNLO LEC, $C_{93}^r$, in good agreement with expectations 
based on dominance by the large $\rho$ peak in $\rho^{I=1}(s)$. 
In the narrow width approximation, expanding the $\rho$ propagator to 
one higher order yields an NNNLO contribution proportional to $Q^4$ 
which is numerically relevant already at $Q^2\sim 0.1$ GeV$^2$. 
We have thus constructed a supplemented ``NN$^\prime$LO'' form by 
adding a term $CQ^4$ to the known NNLO form. Fixing $C_{93}^r$ and 
$C$ to reproduce the first two derivatives of $\hat{\Pi}^{I=1}(Q^2)$ 
with respect to $Q^2$ at $Q^2=0$ yields the representation given by 
the dashed line in Fig.~\ref{fig7}. The dotted line is the analogous 
pure NNLO result. The NNLO form produces $4\%$ and $18\%$ errors
on $\hat{a}_\mu^{\rm LO,HVP}[Q^2_{max}]/a_\mu^{LO,HVP}$ for 
$Q^2_{max}=0.1$ and $0.2$ GeV$^2$, respectively, and is hence 
inadequate for our purposes. The corresponding errors for the 
NN$^\prime$LO form are $0.6\%$ and $1.4\%$, with only the former 
showing sub-percent accuracy. Since the accuracy of even the 
NN$^\prime$LO representation begins to break down above 
$Q^2\sim 0.1$ GeV$^2$, fixing the required LECs via fits to 
$\hat{\Pi}^{I=1}(Q^2)$ on an interval like $0.1\le Q^2\le 0.2$ 
GeV$^2$ is not an option in this case. We thus expect the ChPT 
representation to be less useful than the Pad\'e and 
conformal-variable-polynomial representations, though it will
provide cross-checks on the other methods. One place where
the ChPT representation is, nonetheless, useful, is in providing
an understanding of the dependence of the low-$Q^2$ contributions 
to \hatamulohvp  on $m_\pi^2$ and $m_K^2$. We find that, for
many of the simulations in the literature, $m_\pi^2$ is too large
to allow a linear extrapolation to physical $m_\pi^2$.

\end{document}